\documentclass[aps,prl,twocolumn,showpacs,preprintnumbers,floatfix]{revtex4}

\usepackage{graphicx}
\usepackage{dcolumn}
\usepackage{bm}

\begin{document}

\title{Compound nuclear decay and the liquid to vapor phase transition: a physical picture}

\author{L. G. Moretto, J. B. Elliott and L. Phair}
\affiliation{Nuclear Science Division, Lawrence Berkeley National Laboratory \\
                   University of California, Berkeley, California 94720 }

\date{\today}

\begin{abstract}
Analyses of multifragmentation in terms of the Fisher droplet model (FDM) and the associated construction of a nuclear phase diagram bring forth the problem of the actual existence of the nuclear vapor phase and the meaning of its associated pressure.\ \ We present here a physical picture of fragment production from excited nuclei that solves this problem and establishes the relationship between the FDM and the standard compound nucleus decay rate for rare particles emitted in first-chance decay.\ \ The compound thermal emission picture is formally equivalent to a FDM-like equilibrium description and avoids the problem of the vapor while also explaining the observation of Boltzmann-like distribution of emission times.\ \ In this picture a simple Fermi gas thermometric relation is naturally justified and verified in the fragment yields and time scales.\ \ Low energy compound nucleus fragment yields scale according to the FDM and lead to an estimate of the infinite symmetric nuclear matter critical temperature between $18$ and $27$ MeV depending on the choice of the surface energy coefficient of nuclear matter.
\end{abstract}


\maketitle

After decades of theoretical and experimental studies, recent papers have presented what can be considered a quantitative, credible liquid-vapor phase diagram containing the coexistence line up to the critical temperature for small nuclear systems \cite{elliott-02,elliott-03}.\ \ This diagram was not obtained by ``traditional'' methods of caloric curves \cite{pochodzella-00,chernomoretz-01} or anomalous heat capacities \cite{gulminelli-99,dagostino-00}.\ \ Rather, it was generated from the fitting of the charge distributions in multifragmentation by means of a Coulomb corrected Fisher droplet model (FDM) \cite{elliott-02,elliott-03,fisher-67.1,fisher-67.2} which gives the cluster composition of a vapor:
\begin{equation}
	n_A(T)=q_0A^{-\tau}\exp\left(\frac{\Delta\mu A}{T}-\frac{c_0\varepsilon A^{\sigma}}{T}\right),
\label{eq:Fisher}
\end{equation}
where: $q_0$ is the normalization, $A$ is the cluster mass number, $\tau$ is a topological critical exponent, $c_0$ is the surface energy coefficient, $T$ is the temperature, $\varepsilon=(T_c-T)/T_c$ is the relative deviation from the critical temperature $T_c$, and  $\sigma$ is the surface to volume exponent.\ \ The quantity $c_0 \varepsilon A^{\sigma}$ is the surface free energy cost of forming an $A$-sized cluster.\ \ $\Delta\mu$ is the difference of chemical potentials between the liquid and the vapor.\ \ Finite size effects \cite{moretto-04}, Coulomb energies \cite{moretto-03} and other nuclear energy terms (e.g. due to isospin and angular momentum) can be absorbed into the chemical potential.\ \ In references \cite{elliott-02,elliott-03} this was done for the Coulomb energy by writing $\Delta\mu$ as the Coulomb energy of a residue plus fragment ``saddle'' configuration compared with the initial state.\ \ For $\Delta\mu=0$ the liquid and the vapor are in equilibrium and Eq.~(\ref{eq:Fisher}) can be taken to be the equivalent of the coexistence line \cite{fisher-67.1,fisher-67.2}.

In fact, one can immediately obtain from Eq.~(\ref{eq:Fisher}) the more conventional $p$, $T$ and $\rho$, $T$ phase diagrams by recalling that in the FDM the clusterization is assumed to exhaust all the non-idealities of the gas. It then becomes an ideal gas of clusters.\ \ Consequently, the total pressure and density can be immediately calculated
\begin{equation}
	p(T)=T\sum _A n_A(T),
\label{eq:tot_p}
\end{equation}
\begin{equation}
	\rho(T)=\sum _A An_A(T),
\label{eq:tot_d}
\end{equation} 
as well as the corresponding scaled quantities ${p}/{p_c} $ and ${\rho}/{\rho _c}$

Tests on the lattice gas (Ising) model \cite{moretto-04,mader-03,elliott-04} demonstrate a beautiful agreement between the vapor cluster concentrations and Eq.~(\ref{eq:Fisher}), and analysis of many multifragmentation
reactions \cite{elliott-02,elliott-03} show equally good agreement, leading to a characterization of the liquid-vapor phase diagram for small nuclear systems.

The most troubling point in this otherwise elegant picture is summarized by the question: where is the vapor?\ \ Does the nuclear system truly present itself at some time like a mixed phase system with the vapor being somehow
restrained (as in lattice gas calculations \cite{moretto-04,elliott-04}), either statically or dynamically in contact with the
liquid phase? If so, how is the nuclear vapor restrained?\ \ If not, what is the meaning of vapor pressure, when clearly the system is freely decaying in vacuum against no pressure?

In this paper we present a physical picture of fragment production from excited nuclei that will show: (a) how one can talk about coexistence without the vapor being present; (b) why a simple thermometric equation such as $E=aT^2$ is a more pertinent thermometer for intermediate mass fragment production than empirical thermometers such as isotope thermometers; and  (c) why an equilibrium description, such as the FMD, is relevant to the free vacuum decay of a multifragmenting system.

To begin, let us imagine a liquid in equilibrium with its saturated vapor.\ \ At equilibrium, any particle evaporated by the liquid is restored on the average by the vapor bombarding it.\ \ In other words, the outward evaporation flux from the liquid to the vapor is exactly matched by the inward condensation flux.\ \ This is true for any kind of evaporated particle.\ \ The vapor acts like a mirror, reflecting back into the liquid the particles which it is evaporating.\ \ One obviously can probe the vapor by putting a detector in contact with it.\ \ However, since the outward and inward fluxes are identically the same, one might as well put the detector in contact with the liquid itself.\ \ At equilibrium, the two measured fluxes must be the same.\ \ Therefore, we do not need the vapor to be physically present in order to characterize it completely.\ \ We can just as well study the evaporation of the liquid and dispense with our imaginary surrounding saturated vapor.\ \ The vapor need not be there at all.\ \ One speaks in these situations of a ``virtual vapor'', realizing that first order phase transitions depend exclusively upon the intrinsic properties of the two phases, \emph {and not on their interaction}.\ \ Of course, if the vapor is not there to restore the emitting system with its back flux, evaporation will proceed, leading to a cooling off of the system.\ \ This naturally suggests the study of the emission of fragments from an excited nucleus since the nucleus is a small drop of nuclear liquid evaporating in vacuum; i.e. emitting fragments in vacuum.

We now specifically address fragment emission from an excited nucleus with a time-honored assumption which we do not justify other than through the clarification it brings to the experimental picture: we assume (just as in compound nuclear decay) that, after prompt emission in the initial phase of the collision, the resulting system relaxes in shape and density and thermalizes {\em on a time scale shorter than its thermal decay}.\ \ At this point the excited nucleus emits particles in vacuum, {\em according to standard statistical decay rate theory}.\ \ In this picture there is no surrounding vapor, no confining box, and there no need for either.\ \ By studying the outward flux of the \emph{first} fragments emitted, we can study the nature of the vapor even when it is absent (the {\em virtual} vapor) because of the equivalence of the evaporation and condensation fluxes of a liquid in equilibrium with its saturated vapor.

Quantitatively, the concentration $n_A(T)$ of any species $A$ in the vapor is related to the corresponding decay rate $R_A(T)$ (or to the decay width $\Gamma _A$) from the nucleus by matching the evaporation and condensation fluxes
\begin{equation}
	R_A(T) =\frac{\Gamma_A(T)}{\hbar} \simeq n_A(T) \left< v_A(T) 4 \sigma _{\rm inv}(v_A)\right>,
\label{eq:rate}
\end{equation}
where $v_A(T)$ is the velocity of the species $A$ (of order $(T/A)^{1/2}$) crossing the nuclear interface represented by the cross section $\sigma _{\rm inv}$ (of order $A_0^{2/3}$ where $A_0$ is the mass number of the evaporating nucleus).\ \ The temperature $T_0$ of the equilibrated, excited nucleus when the first fragment is emitted can be estimated by the thermometric equation of a Fermi gas and the calorimetrically mesaured excitation energy $E^*$
\begin{equation}
	T_0 = \sqrt{E^*/a}
\label{fermi-temp}
\end{equation}
allowing for a weak dependence of $a$ on $T$ \cite{hagel-88,raduta-97}, and remembering that the system is most likely still in the Fermi strong degeneracy regime (where the temperature is much less than the Fermi energy: $T \ll \varepsilon_F$).

This is the fundamental and simple connection between Eq.~(\ref{eq:rate}), the (compound nucleus) decay rate, and Eq.~(\ref{eq:Fisher}), the FDM.\ \ In the latter, one immediately recognizes in the exponential the canonical expansion of the standard compound nucleus decay rate, namely the Boltzmann factor $\exp (-B/T)$ where $B$ is the emission barrier which in Eq.~(\ref{eq:Fisher}) is written with its surface factor isolated from all other components, e.g. Coulomb \cite{moretto-03}, symmetry, finite size \cite{moretto-04}, etc.\ \ Thus, the vapor phase in equilibrium can be completely characterized in terms of the decay rate.

The physical picture described above is valid instantaneously, but not globally.\ \ The result of a global evaporation in vacuum leads to abundances of various species of emitted fragments that arise from a continuum of systems at different temperatures.\ \ This leads to complications in various thermometers: kinetic energy, isotope ratios, etc.\ \ One way to avoid this complication is to consider only fragments that are emitted very rarely so that, if they are not emitted first, they are effectively not emitted at all.\ \ In other words, we consider only fragments that by virtue of their large surface energy, have a high emission barrier. 

\begin{figure}
\includegraphics[angle=90,width=8.0cm]{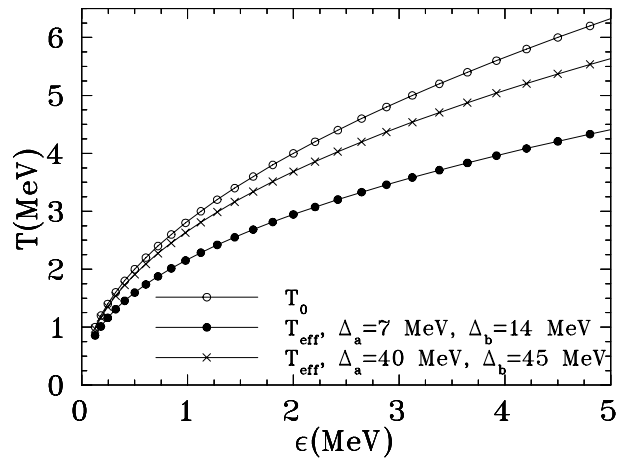}
\caption{\label{fig:cooling} The effective temperature of a Fermi system with three exit channels ($a$, $b$, and $n$) is plotted as function of initial excitation energy for two cases: one where barriers $B_a$ and $B_b$ are large (crosses) compared to $B_n$=6 MeV, and another where $B_a$ or $B_b$ is similar (solid circles) to $B_n$.\ \ The initial temperature as a function of initial excitation energy is shown by the open circles.}
\end{figure}

As an example of the effects of this complication on the isotope thermometer, consider a decaying system with only three available exit channels $a, b, $ and $n$ with barriers $B_a$, $B_b$, and $B_n$. For $B_n \ll B_a$ and $B_b$ we know that the probability of emission of particles of type $b$ at a fixed temperature is approximately
\begin{equation}
	p_b \approx \exp \left[ - \left(B_b-B_n\right)/T \right].
\label{prob}
\end{equation}
Since the nucleus cools as particles are emitted, the total emission probability of particles of type $b$ from a nucleus at $T_0$ is
\begin{equation}
	P_b\propto \int _{0}^{T_0} dT \frac{2aT \exp \left[ - \left(B_b-B_n\right)/T \right]}{B_n + 2 T}.
\end{equation} 
A similar expression holds for $P_a$. The ratio of $P_b/P_a$ is
\begin{equation}
	\frac{P_b}{P_a}\simeq\frac{\Delta _b^2}{\Delta _a^2}
	\frac{\int _0^{T_0/\Delta _b}e^{-1/x}x dx}{\int _0^{T_0/\Delta _a}e^{-1/x}x dx}
\end{equation}
where $\Delta _b=B_b-B_n$ and $\Delta _a=B_a-B_n$. 
The ratio $P_b/P_a$ can also be used to extract an effective temperature $T_{\rm eff}$
\begin{equation}
	{P_b}/{P_a}=\exp\left[- \left({B_b-B_a}\right)/{T_{\rm eff}}\right].
\end{equation}

This sequential aspect is of course lost in equilibrium models which, at best, are corrected for the secondary decay of the primary fragments, but not for the source decay.

A comparison of $T_{\rm eff}$ and $T_0$ is given in Fig.~\ref{fig:cooling} for different values of $B_b$ and $B_a$. The case where $B_a$ and $B_b$ are large (crosses) gives effective temperatures very near to the initial temperature $T_0$ (open circles).\ \ When either $B_a$ or $B_b$ is near the barrier of the most probable channel (solid circles), the effective temperature is very different from the initial temperature. 

Thus, in order to justify the use of the initial Fermi temperatures one should choose exit channels with large barriers.\ \ This was done in the analyses leading to the nuclear phase diagrams \cite{elliott-02,elliott-03}, where
fragments with charge $Z < 5$ were not considered.\ \ Under these conditions, the validity of Eq.~(\ref{eq:rate}) is essentially guaranteed.\ \ The emission rate can then be related to the vapor concentration and the phase diagram can be constructed.\ \ The temperature necessary for our purpose is $T_0$ and not some average temperature determined from multiply emitted particles. 

The correctness of the thermometric relation given by Eq.~(\ref{fermi-temp}) can be tested ``a posteriori'' by verifying the linearity of the Arrhenius plots $\ln n_A$ vs. $1/T$ \cite{moretto-97} and of the socalled Fisher plots of $n_A$ scaled via Eq.~(\ref{eq:Fisher}) \cite{elliott-02,elliott-03}.\ \ This linearity, extending over many orders of magnitude for a variety of fragments, is the strongest test yet of a Fermi gas thermometric relationship.\ \ In fact one can reverse the problem and determine the thermometric relationship up to rather high excitation energies by the requirement that it leads to a linear Arrhenius and Fisher plots.\ \ This is an important, novel insight into the problem of temperature in excited nuclei. 

\begin{figure}
\includegraphics[angle=90,width=8.0cm]{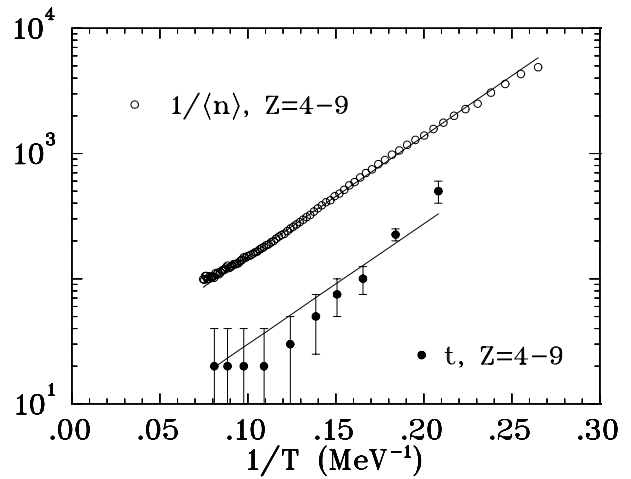}
\caption{\label{fig:mean_time} The mean emission times (in fm/c) of fragments with atomic number $4\le Z\le 9$ are plotted (solid symbols) versus inverse temperature for the reaction $\pi$+Au at 8 GeV/c \protect\cite{beaulieu-01a,beaulieu-01b}. The average yields  of the same fragments are plotted versus $1/T$ (solid symbols). The line represents a Boltzmann fit to the fragment yields. This same line has been superimposed (shifted) on to the emission times.}
\end{figure}

Further evidence for our physical picture of an excited nucleus evaporating fragments can be found in the mean emission time of fragments.\ \ To see this, we first construct an Arrhenius plot ($n_A(T)$ versus $1/T$) of the abundances of observed fragments as a function temperature such as the one in Fig.~\ref{fig:mean_time} which shows the reciprocal of the average abundances $1 / \left< n \right>$ for fragments with $4 \le Z \le 9$ measured by the ISiS collaboration in the reaction $\pi$+Au at 8 GeV/c  \cite{beaulieu-01a,beaulieu-01b}.

The slope of the Arrhenius plot is the effective ``barrier'' $B_A$ for the emission of the fragment.\ \ This is evident in Eq.~(\ref{eq:rate}) which gives
\begin{equation}
	n_A(T) \propto \Gamma_A(T) \propto \exp(-B_A / T).
\label{yields-0}
\end{equation}
Equation~(\ref{eq:rate}) also shows that the same barrier and the same Boltzmann factor determine the mean emission time $t$ of a fragment since
\begin{equation}
	t_A(T)=\frac{\hbar}{\Gamma_A(T)}\propto \exp(B_A / T).
\label{time}
\end{equation}
Such a time $t_A(T)$ is the reciprocal of $\Gamma_A(T)$.

Therefore, the same Arrhenius plot with the same barrier should describe both the temperature dependence of the abundances and of the times.\ \ This is exactly the case shown in Fig.~\ref{fig:mean_time}.\ \ The ISiS collaboration has measured the yields (open symbols) \cite{elliott-02} and the mean emission times $t$ (solid symbols) \cite{beaulieu-01a,beaulieu-01b} of fragments with $4 \le Z \le 9$ as a function of excitation energy.\ \ These energies can be translated into a Fermi gas temperature \cite{elliott-02} as discussed above.\ \ A Boltzmann fit to the yields is shown by the solid line.\ \ That same line has been superimposed (shifted) onto the emission time data which it describes very well.\ \ Interestingly, the two different observables and their energy dependence are described by equations (\ref{yields-0}) and (\ref{time}) with the same barrier.

All that has been written above holds exactly for low excitation energies, thus ordinary compound nuclear decay is directly relevant to the liquid-vapor phase transition.\ \ We can then find further evidence for our physical picture by scaling known low energy compound nucleus fragment yields \cite{fan-00} according to the FDM.

In compound nucleus (and multifragmentation) experiments we measure the fragment yields $Y_A(T)$ (the number of fragments in events with a given excitation energy or temperature divided by the total number of events at that excitation energy or temperature) rather than the fragment concentrations $n_A(T)$ (the number of fragments of a given mass per unit volume).\ \ However, following the logic set forth above, we can relate the measured yields to the equilibrium fragment concentrations as follows: since the fragments in question are large ($Z \ge 7$) they are emitted from the compound nucleus of $A_0$ nucleons first or not at all.\ \ Then for first chance fragments the yield is given by:
\begin{equation}
 Y_A(T) = \frac{\Delta tR_A(T)}{\Delta tR_n(T) + \Delta t \sum_{A\ne n}R_A(T) }
\label{yields-1}
\end{equation}
where $\Delta t$ is the time duration of the measurement and $R_n(T)$ is the rate of neutron decay rate.\ \ Since neutron emission is much more probable than heavy fragment emission, in other words $R_n(T) \gg  \sum_{A \ne n}R_A(T)$, the yield of a given fragment is 
\begin{equation}
 Y_A(T)  \simeq \frac{R_A(T)}{R_n(T)}
\label{yields}
\end{equation}
Recalling that
\begin{equation}
R_n(T) \simeq \frac{T}{\hbar} \exp \left( -\frac{B_n}{T} \right)
\label{frag-prob}
\end{equation}
where $B_n \approx 8$ MeV is the neutron binding energy.\ \ It follows from equations (\ref{eq:rate}), (\ref{yields}) and (\ref{frag-prob}) that
\begin{eqnarray}
 Y_A(T) &  \simeq & \frac{\hbar}{T} \exp \left( \frac{B_n}{T} \right) n_A(T) \left< v_A(T) 4 \sigma _{\rm inv}(v_A)\right> \nonumber \\
 & \simeq & {q_0}^{\prime} \frac{\hbar A_0^{\frac{2}{3}} }{\sqrt{AT}} A^{-\tau} \exp\left(\frac{B_n+\Delta\mu-c_0\varepsilon A^{\sigma}}{T}\right)
\label{yields-full}
\end{eqnarray}
where ${q_0}^{\prime}$ consolidates the constants from $v_A(T)$, $4 \sigma_{\rm inv}(V_A)$ and $n_A(T)$.\ \ The measured first chance fragment yields are thus proportional to the concentration of the virtual vapor.

\begin{figure}
\includegraphics[width=8.7cm]{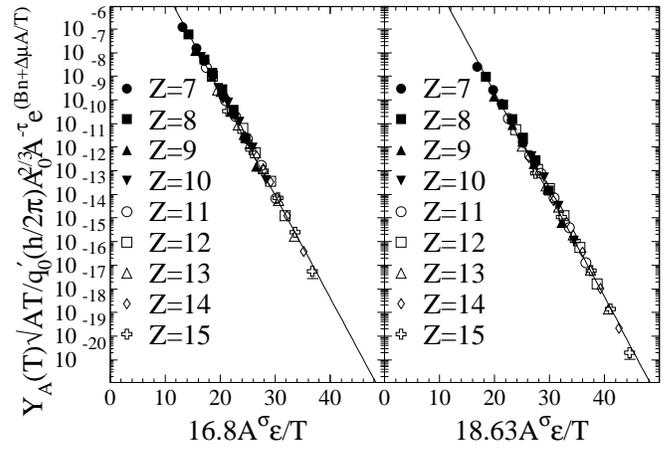}
\caption{\label{fig:cn} Results for the FDM-scaled yield distribution for the $^{64}$Ni$+^{12}$C compound nucleus decay data.\ \ See text for details.}
\end{figure}

\begin{table}[htdp]
\caption{Results for fitting parameters}
\begin{center}
\begin{tabular}{ccc}
\hline
        $c_0$                     & $16.8$ MeV          & $18.63$ MeV        \\
\hline
        ${\chi}^{2}_{\nu}$ &$1.39$                    & $1.53$                    \\
        ${q_0}^{\prime} $ &$40\pm20$     & $300\pm 100$ \\
        $\Delta \mu$         &$2.43\pm0.05$MeV & $3.01\pm0.05$ MeV \\
        $T_c$                     &$10.6\pm0.6$MeV      & $14\pm1$ MeV     \\
\hline
\end{tabular}
\end{center}
\label{results}
\end{table}

The yields of charged fragments from the reaction $^{64}$Ni+$^{12}$C \cite{fan-00} were fit to Eq.~(\ref{yields}).\ \ These data were taken at the Berkeley 88-inch cyclotron using Ni beams with energies between $6$ and $13$ AMeV suggesting $A_0=76$, verified experimentally \cite{fan-00} and excitation energies of $0.96$ AMeV $\le E^* \le 1.82$ AMeV and temperatures (calculated as in references \cite{elliott-02,elliott-03,raduta-97}) of $2.9$ MeV $\le T \le 4.2$ MeV.\ \ These excitation energies are small and the fragment emission barriers are large compared to those of neutron evaporation; therefore there is little doubt about the validity of the thermometric relation of Eq.~(\ref{fermi-temp}).\ \ The mass of a fragment $A$ was estimated from the measured charge of a fragment $Z$ using the EPAX parameterization \cite{summerer-90}.\ \ Because the excitation energies are low, the effects of the secondary decay of the fragments is minimal.

$54$ points for fragments with charges of $7 \le Z \le 15$ from the $^{64}$Ni+$^{12}$C reaction were fit to Eq.~(\ref{yields}) with three free parameters: ${q_0}^{\prime}$, $\Delta \mu$ and $T_c$.\ \ The critical exponents $\tau=2.209\pm0.006$ and $\sigma=0.63946\pm0.0008$ were set to their standard three dimensional Ising values \cite{mader-03}.\ \ As a test of the systematic errors the surface energy coefficient $c_0$ was set to its text book value $16.8$ MeV \cite{krane-text} from the Weiz\"{a}cker-Bethe semiempirical formula and to $18.63$ as suggested by more recent efforts \cite{myers-01}.

Results are shown in Fig.~\ref{fig:cn} and given in Table~\ref{results}.\ \ The meaning of the value of the normalization ${q_0}^{\prime}$ is unclear, though it must depend on the constants or proportionality in $v_A(T)$, $\sigma_{\rm inv}(V_A)$ and $n_A(T)$.\ \ The {\em effective} chemical potential $\Delta \mu$ value is due to the effects of finite size \cite{moretto-04}, the energetic contributions from Coulomb \cite{moretto-03} and other nuclear effects (e.g. isospin, angular momentum, etc.) and the Coulombic contributions to $\sigma_{\rm inv}$.\ \ It is assumed that $\Delta \mu$ is approximately constant for the $76$ nucleon system over the small temperature range considered.\ \ The value of the critical temperature between $10.6 \pm 0.6$ and $14 \pm 1$ MeV (whose value is essentially dictated by the choice of the surface energy coefficient of nuclear matter) can be interpreted as the critical temperature of bulk, neutral, symmetric nuclear matter since the {\em effective} chemical potential $\Delta \mu$ absorbs all but surface effects.

\begin{figure}
\includegraphics[width=8.7cm]{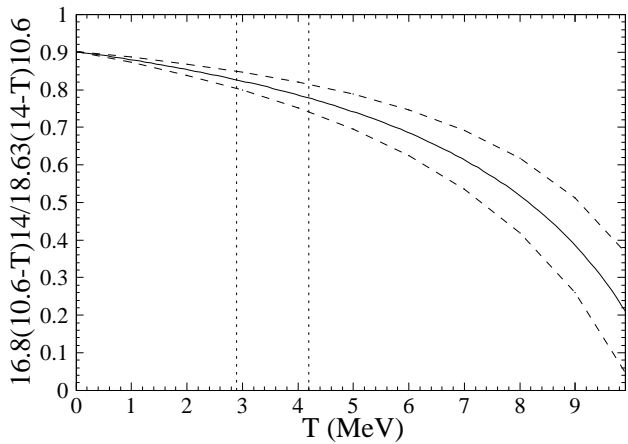}
\caption{The solid line shows the ratio of $c_0\varepsilon$ (the surface free energy coefficient) for the two estimates of the nuclear binding energy used as a function of temperature.\ \ Dashed lines show the error arising from the errors on the $T_c$ values returned by the fitting procedure.\ \ Dotted vertical lines show the temperature range covered by the compound nucleus experiment.}
\label{range}
\end{figure}

The systematic error of the fit parameters in this analysis is sizable.\ \  This is not unexpected since the range in temperature covered by the compound nucleus experiment is small compared to either value of $T_c$ thus prviding a limited range over which to test the dependence of the fragment yieds on $c_0 \varepsilon$ (the surface free energy coefficient).\ \ The solid line in Fig.~\ref{range} shows the ratio of $c_0 \varepsilon$ for the two estimates of the nuclear binding energy used as a function of temperature, the dotted vertical lines showing the temperature range covered by the compound nucleus experiment.\ \ The value of this ratio is nearly constant for the temperatures considered here.\ \ The difference of the ratio from unity is accounted for by the differing values of ${q_0}^{\prime}$ and $\Delta \mu$.\ \ Without a larger lever arm in temperature and a better understanding of ${q_0}^{\prime}$ and $\Delta \mu$ this level of systematic error is unavoidable.

In conclusion, we have shown that: a) first chance emission from a hot thermalized source is a natural explanation for the scaling of experimentally observed intermediate mass fragment yields according to Eq~(\ref{eq:Fisher}); b) The simple relationship between emission rates and saturated vapor concentration neatly eliminates the unphysical assumption of a physical vapor in equilibrium with a liquid; c) Fermi gas like temperatures associated with first chance emission find their natural explanation and are justified by scaling plots such as Fig.~\ref{fig:cn}; d) The emission rate picture explains the identical slopes  in the Arrhenius plots for the mean multiplicities and the mean emission times shown in Fig.~\ref{fig:mean_time}; e) Finally, the fragment emission from an experimentally certified low energy compound nucleus reaction scales according to Eq.~(\ref{eq:Fisher}) and leads among other things to a value of the infinite nuclear matter critical temperature between $10.6 \pm 0.6$ and $14 \pm 1$ MeV in agreement with theoretical estimates \cite{friedmann-81,jaqaman-83,jaqaman-84,levit-85,glendenning-86,mueller-95,de-99}.

\begin{acknowledgments}
This work was supported by the US Department of Energy.
\end{acknowledgments}

\end{document}